# Accelerating Cosmological Models for an Interacting Tachyon

U. Filobello[1] and A. de la Macorra[2,3]

[1]Facultad de Instrumentación Electrónica, Universidad Vercacruzana,
Apdo. Postal 465, Xalapa Veracruz, México.

[2]Instituto de Física, Universidad Nacional Autónoma de México,
Apdo. Postal 20-364, 01000 México D.F., México.
[3]Parto of the Collaboration Instituto Avanzado de Cosmologia

We study the cosmological evolution of a tachyon scalar field T with a Dirac- Born- Infeld type lagrangian and potential $V(T)$ coupled to a canonically normalized scalar field $\phi$ with an arbitrary but factorizeable interaction potential B (T, $\phi$) in the presence of a barotropic fluid. We will assume that the force between the barotropic fluid and the scalar field is only gravitational and we show that all models dependence is given in terms of three parameters: $\lambda_1 = -V_T/V$, $\lambda_2 = -B_T/B$ and $\lambda_3 = -B_\phi/B$, which are field and time dependent. We give the general solution to the cosmological equations of motion and determine in which cases we can have an accelerating universe giving rise to dark energy.

## I INTRODUCTION

Various observational [1] evidences including SN1a data leads to one of the most important result in modern cosmology; we live in a universe where the energy density is dominated by dark energy "DE" (an almost uniform distribution of energy density with a negative pressure which dominates all other form of matter at present time), it means that the universe not only expanding but it is accelerating. The current dates give a flat universe with $\Omega_{DM} = 0.22$, $\Omega_{BM} = 0.04$ and a dark energy $\Omega_{DE} = 0.74$ with an equation of state in the range $-1.11 < w < -0.86$ [1].

Instead of dealing directly with a static cosmological constant perhaps a better idea is to dynamical generating DE via the existence of scalar fields with the goal to help us to understand the nature of the acceleration of the universe. There is a large class of scalar field models with a large variety of scalar potentials or different kinetic terms such as canonical or tachyonic fields. Scalar fields, canonical or tachyonic, have been widely studied in the literature. Canonical scalar fields are obtained in particle physics and they can be either fundamental or composite fields and can be used to describe the dark energy as quintessence [7,12]. Tachyon fields are originated from the D-brane action (Dirac-Born-Infeld type lagrangian) in string theory and they are the lowest energy state in unstable Dp-brane systems [2]. As they represent the low energy limit of string brane models its phenomenology is important and a great amount of work has been invested in studying the dynamics of tachyon field [2], [3], [10].

One of the main goals is to find models which leads asymptotically to an accelerating universe leading to the presence of dark energy [3],[9],[10]. In this letter we study the cosmological evolution of a tachyon field T coupled to a canonical scalar field $\phi$ through an arbitrary but factorizeable interaction potential B (T, $\phi$) = F(T)$G(\phi)$ in the presence of a barotropic fluid, that can be either matter or radiation, and we will show that all models dependence is given in terms of three parameters $\lambda_1 = -V_T/V$, $\lambda_2 = -B_T/B$ and $\lambda_3 = -B_\phi/B$ which in general evolve in time.

Here, we generalize a previous work [6] where we assumed constant parameters $\lambda_i's$. The case of constant $\lambda_i's$ is indeed restricted (because the cases $\lambda_1 =$ constant and $\lambda_3 =$ constant correspond to the specific scalar potentials $V(T) = AT^{-2}$ and $B(T,\phi) = AF(T)e^{-c\phi}$ respectively). Our analysis here is completely general and model independent, we do not assume any kind of tachyon potential $V(T)$. The dynamical equations of motion are none linear and general analytic solutions do not exist. We determined the critical points for the cases where the universe ends accelerating and it give rise to twenty two possible models. To study them result natural consider separately the cases: $\lambda_3 = cte \neq 0$, $\lambda_3 = 0$ and $\lambda_3 = \infty$. At the same time we will emphasize the convenience of separate the models where $\lambda_2 \to 0$ from the models where $\lambda_2 \neq 0$ asymptotically.

The paper is organized as follows: In section II we give the dynamical differential equations of the system and we rewrite them in an autonomous form. In section IIA we present a review of the relevant results obtained previously [6]. In section III we give a detailed description of the models emphasizing the fact that they were selected to result asymptotically in an accelerating universe (with an equation of state parameter $w_{effec} = -1$ for the dominant fluid). Finally, we summarize and conclude in section IV.

## II COUPLED TACHYON AND SCALAR FIELD

We will study the dynamics for a system of two interacting scalar fields T, $\phi$ where the tachyon field $T$ is given by a Dirac-Born-Infeld type lagrangian with a potential $V(T)$ and $\phi$ represent a standard canonical field. We also incorporate a barotropic energy density $\rho_b$, which can be either matter $w_b = 0$ or radiation $w_b = 1/3$. We will assume that the scalar fields interact via a potential $B(T,\phi)$, while there is only gravitational interaction between these fields and the barotropic fluid. We take the following Lagrangian for the scalar fields T and $\phi$ [6].

$$L = -V(T)\sqrt{1-\partial_\mu T\partial^\mu T} + \frac{1}{2}\partial_\mu\phi\partial^\mu\phi - B(T,\phi) \qquad (1)$$

The equations of motion of $\phi$ and $T$ for a spatially flat Friedman-Robertson-Walker universe are

$$\ddot{\phi} + 3H\dot{\phi} + B_\phi = 0\,, \qquad (2)$$

$$\frac{\ddot{T}}{\sqrt{1-\dot{T}^2}} + 3H\dot{T} + \frac{V_T}{V} + B_T\frac{\sqrt{1-\dot{T}^2}}{V} = 0 \qquad (3)$$

where the subindex in V and B is defined as $V_T \equiv \partial V/\partial T$, $B_\phi \equiv \partial B/\partial\phi$ and $B_T \equiv \partial B/\partial T$. The Hubble parameter $H = \dot{a}/a$ is

$$3H^2 = \rho = \rho_\phi + \rho_T + \rho_b\,, \qquad (4)$$

where we have taken $8\pi G \equiv 1$, $\rho$ is the total energy density, $\rho_b$ the barotropic fluid and $\rho_T, \rho_\varphi$ are the defined as

$$\rho_\phi \equiv \frac{1}{2}\dot{\phi}^2 + B(T,\phi)\,,\; P_\phi \equiv \frac{1}{2}\dot{\phi}^2 - B(T,\phi) \qquad (5)$$

for $\phi$ and

$$\rho_T \equiv \frac{V(T)}{\sqrt{1-\dot{T}^2}}\,,\; P_T \equiv -V(T)\sqrt{1-\dot{T}^2} \qquad (6)$$

for T with $P_T, P_\phi$ the pressure of $T,\phi$ respectively. The ratio of densities are $\Omega_T \equiv \rho_T / 3H^2$, $\Omega_\phi \equiv \rho_\phi / 3H^2$, $\Omega_b \equiv \rho_b / 3H^2$. Using eqs. (5) and (6) we can rewrite the dynamical eqs. (2) and (3) in terms of the energy densities as

$$\dot{\rho}_\phi + 3H\rho_\phi\left(1+w_\phi\right) = B_T\dot{T} = \delta ,$$
$$\dot{\rho}_T + 3H\rho_T\left(1+w_T\right) = -B_T\dot{T} = -\delta , \tag{7}$$
$$\dot{\rho}_b + 3H\rho_b\left(1+w_b\right) = 0 ,$$

where we have included the evolution of the barotropic fluid $\rho_b$ and

$$\delta \equiv B_T\dot{T} , \tag{8}$$

defines the interaction term. The equation of state parameters are given by

$$w_\phi \equiv \frac{P_\phi}{\rho_\phi} = \frac{\dot{\phi}^2/2 - B(T,\phi)}{\dot{\phi}^2/2 + B(T,\phi)}, \qquad w_T \equiv \frac{P_T}{\rho_T} = -1 + \dot{T}^2 , \qquad w_b \equiv \frac{P_b}{\rho_b} . \tag{9}$$

In order to have a real energy density for the tachyon we require $0 \le \dot{T}^2 \le 1$ and from eq. (9) we see the equation of state parameter for T is constrained to $-1 \le w_T \le 0$. In order to definean equation effective equation of state it is convenient to rewrite eqs. (7) as:

$$\dot{\rho}_\phi + 3H\rho_\phi\left(1+w_{\phi eff}\right) = 0 , \qquad \dot{\rho}_T + 3H\rho_T\left(1+w_{Teff}\right) = 0 , \tag{10}$$

with the effective equation of state defined by

$$w_{\phi eff} \equiv w_\phi - \frac{B_T\dot{T}}{3H\rho_\phi}, \qquad w_{Teff} \equiv w_T + \frac{B_T\dot{T}}{3H\rho_T} . \tag{11}$$

To determine the attractor solutions of the differential equations given in eqs. (2), (3) and (4) it is useful to make the following change of variables:

$$X_1 = \dot{T} = HT', \qquad Y_1 = \frac{\sqrt{V}}{\sqrt{3}H}, \tag{12}$$

$$X_2 = \frac{\dot{\phi}}{\sqrt{6}H}, \qquad Y_2 = \frac{\sqrt{B}}{\sqrt{3}H} \tag{13}$$

in such a way that the mentioned system become a set of dynamical differential equations of first order

$$X_1' = -\left(1-X_1^2\right)\left(3X_1 - \sqrt{3}Y_1\lambda_1 - \sqrt{3}\sqrt{1-X_1^2}\,\frac{Y_2^3}{Y_1^2}\lambda_2\right), \tag{14}$$

$$X_2' = -\left(3+\frac{H'}{H}\right)X_2 + \sqrt{\frac{3}{2}}Y_2^2\lambda_3 , \tag{15}$$

$$Y_1' = -\frac{H'}{H}Y_1 - \frac{\sqrt{3}}{2}\lambda_1 X_1 Y_1^2 , \tag{16}$$

$$Y_2' = -\frac{H'}{H}Y_2 - \frac{\sqrt{3}}{2}\lambda_2 X_1 Y_2^2 - \sqrt{\frac{3}{2}}\lambda_3 X_2 Y_2 ,, \tag{17}$$

$$X_1 = 0, \tag{18}$$

$$X_2 = 0, \tag{19}$$

where a prime denotes derivative respect to logarithm of the scale factor " $a$ ", $N \equiv Lna$ , and $H'/H$ is given by

$$\frac{H'}{H} = -\frac{3}{2}[2X_2^2 + X_1^2\Omega_1 + \Omega_b(1+w_b)]. \tag{20}$$

We have defined the parameters:

$$\lambda_1 = -\frac{V_T}{V^{3/2}} , \quad \lambda_2 = -\frac{B_T}{B^{3/2}} , \; \lambda_3 = -\frac{B_\phi}{B} . \tag{21}$$

Notice that all model dependence in eqs. (14) to (19) is through the three quantities $\lambda_i(N), i = 1,2,3$ and the constant parameter $\omega_b$ . We note that the mentioned system of equations is closed even for the general case of $\lambda_i(N)$ changing in time. In terms of this variables the acceleration of the universe is given by

$$\frac{\ddot{a}}{a} = H^2\left(1+\frac{H'}{H}\right) = -\frac{H^2}{2}\left(\Omega_b\left(1+3w_b\right)+\Omega_1\left(3X_1^2-2\right)+4X_2^2-2Y_2^2\right), \tag{22}$$

the condition to have $\ddot{a}$ >0, from the last equation is

$$Y_2^2 + \Omega_1\left(1-3X_1^2/2\right) > 1/2\,\Omega_b\left(1+3w_b\right)+2X_2^2 . \tag{23}$$

The flatness condition now reads

$$1 = \frac{\rho_b}{3H^2} + \frac{Y_1^2}{\sqrt{1-X_1^2}} + X_2^2 + Y_2^2 , \tag{24}$$

where we used the ratios of densities $\Omega_1$ and $\Omega_2$ can be expressed in terms of these variables as

$$\Omega_1 = \frac{Y_1^2}{\sqrt{1-X_1^2}} \tag{25}$$

$$\Omega_2 = X_2^2 + Y_2^2 . \tag{26}$$

The effective equations of state defined by eq. (11) are given by

$$w_{1eff} \equiv w_{Teff} = -1 + X_1^2 - \frac{\lambda_2 X_1 Y_2^3}{\sqrt{3}\Omega_1} , \tag{27}$$

$$w_{2eff} \equiv w_{\phi eff} = \frac{X_2^2 - Y_2^2}{X_2^2 + Y_2^2} + \frac{\lambda_2 X_1 Y_2^3}{\sqrt{3}\Omega_2} , \tag{28}$$

where we used the notation $w_1 = w_T$ , $w_2 = w_\phi$ . Finally we deduce an alternative expression for eq. (27) as follows: from the condition for fixed point $X_1' = 0$ we deduce that $X_1\lambda_2Y_2^3/\sqrt{3}\,\Omega_1 = X_1^2 - Y_1\lambda_1X_1/\sqrt{3}$ , in such a way that combining it with eq. (27) we get:

$$w_{1efec} = -1 + \lambda_1 Y_1 X_1/\sqrt{3} . \tag{29}$$

## II A. A REVIEW OF MODELS WITH $\lambda_2 = 0$

Our main goal in this work is to find models which leads asymptotically to an accelerating universe leading to the presence of dark energy. We have said that all models dependence is through the three parameters $\lambda_1 = -V_T/V^{3/2}$ , $\lambda_2 = -B_T/B^{3/2}$ and $\lambda_3 = -B_\phi/B$ which in general evolve in time. One case particularly simple is when $\lambda_2 = 0$ because we get two uncoupled scalar fields with a vanishing interaction term. In this case, it is possible to deduce the solutions from the combination of the works given in [7], [8] where a canonically normalized scalar field or tachyon field in the presence of a barotropic fluid are studied. In ref. [7] it was shown that all models dependence is given in terms of just one parameter $\lambda$,which is given by $\lambda = -V_\phi / V$ (where $\phi$ denotes the canonical scalar field, $V(\phi)$ is the scalar potential, and $V_\phi = dV/d\phi$) and it is identical to our $\lambda_3$. In the same way from the reference [8] the cosmology of a tachyon field in presence of a barotropic fluid, is just determined through the parameter $\lambda = -V_T / V^{3/2}$ (where $T$ denotes the tachyon, $V(T)$ is the tachyon potential and $V_T = dV/dT$ ) which is our parameter $\lambda_1$. From these considerations is worth to present the relevant aspects of the accelerating models in the case where $\lambda_2$ is identically zero, which is equivalent to having no interaction term between both scalar fields. This analysis can help us to understand the cosmological evolution of our models in the case where asymptotically $\lambda_2 \to 0$.

We will base our discussion in the already mentioned references: [7] and [8] (however when we explain the cases of $\lambda = c$ and $\lambda = \infty$ for the tachyon field we will also use too the references [9] and [10] respectively). From [7] we know that any scalar potential $V(\phi)$ leads to one of the three different limiting cases of $\lambda$: finite constant, zero or infinity. In the first case the scalar potential has an exponential form, $V = e^{-c\phi}$ with $\lambda = c$ constant. The EoS parameter is given by $w = -1 + \lambda^2/3$ an it is possible to have accelerating models depending on the value of $\lambda$, which requires $\lambda^2 = c^2 < 2$. The second case is $\lambda \to 0$ and in this limit the first derivative of the potential approaches zero faster than the potential itself and examples of this kind of behavior are given by potentials of the form $V = V_0 \phi^{-n}, n > 0$. In this case the scalar field will dominate the energy and accelerate the universe with an EoS $w = -1$ [7]. To study the limit of $\lambda \to \infty$ we will separate it into two different cases. The first is when $\lambda$ approaches its limiting value without oscillating and the later is when $\lambda$ do oscillate. In the first case we can show that at late times the EoS $w$ approches that of the background dominant energy density but with $w \geq w_b$ . Examples of this kind of behavior are given by potentials like $V = e^{-\alpha\phi^2}$ and $V = e^{-\alpha e^{\phi}}$. When $|\lambda| \to \infty$ but with an oscillating $\phi$ field, we consider the potentials that can be expressed as a power series in $\phi$ and keeping the leading term we have $V = V_0 \phi^n, n > 0$ and even, since the potential must be from below, giving an EoS $w = -1 + 2n/(n+2) = (n-2)/(n+2)$, so for $n = 2$ one has $w = 0$ and $n = 4$ $w = 1/3$. Clearly these models are not good candidates for quintessence.

We know that any tachyon potential $V(T)$ leads to one of the three different limiting cases of $\lambda$: finite constant, zero or infinity [8]. In the first case the tachyon potential corresponds to the inverse square potential $V(T) = AT^{-2}$ and it is possible to have accelerating models depending on the value of $A$. From [9] we know that we get an asymptotic accelerating universe only if $A \geq 2/\sqrt{3}$. The second case is when $\lambda$ approaches zero dynamically. We can classify the situation in two classes [8]: $\lambda \to 0$ without any oscillations of $T$, and $\lambda \to 0$ with $T$ oscillating. In the first case we have potentials as $V(T) = AT^{-n}$ with $0 < n < 2$, and $V(T) = Ae^{1/BT}$ and in this class of models the tachyon field evolves to $T \to \infty$ without oscillations and the universe exhibits acceleration at late times. This is a dark energy scenario in which the future universe is dominated by the tachyon with an equation of state $w_T = -1$. In the second case where $T$ oscillates, an example exhibiting this kind of behavior is the potential $V(T) = Ae^{T^2}$. Since the potential has energy $A$ at the minimum of the potential (T = 0) this eventually leads to an acceleration of the universe while the field oscillates. Finally we consider the case of $|\lambda| \to \infty$. One can classify the tachyon potentials into two classes: $|\lambda| \to \infty$ without the oscillation of the field $T$ and $|\lambda| \to \infty$ with the oscillation of the field. One example of the first class of potentials is given by $V(T) = Ae^{-\beta T}$ with $\beta$ positive and in this case we don't get an asymptotic accelerating universe, however from [8] we find that a transient acceleration can occur in the region where $\lambda$ is smaller than the order of the unity. When $T$ oscillates around the minimum of the potential, (i.e. $V(T) = V_0(T - T_0)^n$, with $n > 0$ and even) the dynamics of T corresponds to the (time-averaged) equation of state. In [10] it was shown that it corresponds to $\langle P_T / \rho_T \rangle = -1/n+1$ ($n > 0$ and even). Since $w_T < w_\gamma$, then the tachyon ends dominating the universe but we don't expect to have an scenario of dark energy because for example, from [8] we know that the equation of state of the tachyon has to be in the interval $w_T \in [-1, -1/3)$ if it has to lead to an accelerating universe while clearly $\langle P_T / \rho_T \rangle \in [-1/3, 0)$.

## III ANALYSIS OF MODELS

In this paper we focus our attention in finding models which leads asymptotically to an accelerating universe. With the end to find them in a systematic way we observe from the eq. (20) that if $X_1^2 = 0$, $X_2^2 = 0$, $\Omega_b = 0$ asymptotically, then $\dot{H} = H'/H = 0$ and from eq. (22) we get $\ddot{a}/a = H^2 + \dot{H} = H^2$, that is an accelerating universe, and $Y_1^2 + Y_2^2 = 1$ (see eq. (24)). We will consider the more general case where some or all of the $\lambda_i$ ($i = 1, 2, 3$) parameters can growth asymptotically to infinite. The solution to (14) to (19) requires the constraints:

$$X_1 = X_2 = 0 \tag{30}$$

$$\lambda_1 Y_1 = -\lambda_2 Y_2^3 / Y_1^2 \quad \to \quad V_T + B_T = 0, \tag{31}$$

$$\lambda_3 Y_2^2 = 0, \tag{32}$$

$$\lambda_1 X_1 Y_1^2 = 0, \tag{33}$$

and

$$\lambda_2 X_1 Y_2^2 = -\sqrt{2}\lambda_3 X_2 Y_2 \ . \tag{34}$$

For the case of a factorizeable interaction potential $B(T,\phi) = F(T)G(\phi)$, eq. (31) acquires the form

$$G(\phi) = -V_T / F_T \ . \tag{35}$$

Since the dynamics of the scalar fields $\phi$ and $T$ is to minimize the potentials $G$ and $V + B$ respectively, then eq.(35) gives us the asymptotic relation between the evolutions of the scalar fields $T$ and $\phi$, in order to reach the mathematical conditions: $V_T + B_T = 0$ and $G_\phi = 0$ of eqs. (31) and (32). Of course, in order to have a non trivial solution to eq.(31) the derivatives B_T and V_T must have different signs, i.e. $-\lambda_2 / \lambda_1 > 0$, since we take in all cases $Y_1$ and $Y_2$ non negative.

Let us first discuss the case when $\lambda_2/\lambda_1 > 0$. In this case the equations that need to be satisfied are $\lambda_1 Y_1 = \lambda_2 Y_2^3/Y_1^2 = \lambda_3 Y_2^3 = 0$, as seen from eq.(31) and (32) with $X_1 = X_2 = 0$ and eqs.(34), eq.(35) is trivally solved. In order to have $Y_2 \neq 0$ we need $\lambda_2 = \lambda_3 = 0$ while $Y_1$ can be different from zero only if $\lambda_1 = 0$. So if $\lambda_1 = 0$ and $\lambda_2$ or $\lambda_3$ are different than zero one has $Y_1 = 1$ and $Y_2 = 0$, since $Y_1^2 + Y_2^2 = 1$, while for $\lambda_1 \neq 0$ and $\lambda_2 = \lambda_3 = 0$ we have $Y_1 = 0$ and $Y_2 = 1$. In table I we summarize the possible models. From eq.(31), (33) and (29) we see that in all cases we have $w_{1eff}$ = -1 but when $Y_1 = 0$ we must have $w_{1eff} = w_{2eff} = -1$. When $\lambda_2 \neq \infty$ then $w_{2eff} = w_2$, i.e. the interacation term in eq.(28) vanishes, but when $\lambda_2 = \infty$ the value of $w_{2eff}$ is model dependent with the constraint $w_{2eff} \geq w_{1eff}$ and the interaction term is not necessarily zero. In the limit $X_2 \to 0$ and $Y_2 \to 0$ the ratio $X_2/Y_2 = C$ is not determine and $w_2 = (Y_2^2 - X_2^2)/(Y_2^2 + X_2^2) = (C^2 - 1)/(C^2 + 1)$ with $0 \leq C^2 \leq \infty$ giving $-1 \leq w_2 \leq 1$. In case that $\lambda_3$ is constant we can anticiptae the value of $w_2$ from the discusion in the previous section, wheren the interaction term vansihes. On the other hand for $X_1 \to 0$ we have $w_1 = -1$ independetly of the value of $Y_1$.

| Num | $\lambda_1$ | $\lambda_2$ | $\lambda_3$ | $Y_1^2$ | $Y_2^2$ | $w_{1eff}$ | $w_{2eff}$ |
|---|---|---|---|---|---|---|---|
| 1 | $\lambda_1$ | $0$ | $\neq 0$ | $1$ | $0$ | $-1$ | $w_2$ |
| 2 | $\lambda_1$ | cte | $\lambda_3$ | $1$ | $0$ | $-1$ | $w_2$ |
| 3 | $\lambda_1$ | $\infty$ | $\lambda_3$ | $1$ | $0$ | $-1$ | $w_{2eff}$ |
| 4 | $\neq 0$ | $0$ | $0$ | $0$ | $1$ | $-1$ | $-1$ |
| 5 | $0$ | $0$ | $0$ | $Y_1^2$ | $1$ - $Y_1^2$ | $-1$ | $-1$ |

Table I. In this table we shows the different models that lead to an accelerating universe with $Y_1^2 + Y_2^2 = 1$ and when $\lambda_2/\lambda_1 > 0$. The values of $\lambda_1, \lambda_3$ are aribitray and can take any value (0, cte or $\infty$) while $w_{2eff}$ in case 3 is model dependent.

In the case where A= $-\lambda_2/\lambda_1 > 0$, then eq.(31) can be solved non trivally and we get the simple solution $(Y_1/Y_2)^3 = -\lambda_2/\lambda_1$ which togther with the condition $Y_1^2 + Y_2^2 = 1$ gives $Y_1^2 = \left(1 + \left(-\lambda_1/\lambda_2\right)^{2/3}\right)^{-1}$ and $Y_2^2 = 1 - \left(1 + \left(-\lambda_1/\lambda_2\right)^{2/3}\right)^{-1}$ or equivalently

$Y_1{}^2 = 1-(1+(-\lambda_2/\lambda_1)^{2/3})^{-1}$ and $Y_2{}^2 = (1+(-\lambda_2/\lambda_1)^{2/3})^{-1}$. The models that solve eqs.(30)-(34) are given in Table II and in the last column we give the constraint on the limiting value of $\lambda_2/\lambda_1$. In case from eq.(32), we see that if $\lambda_3 = 0$ we must have $Y_2 = 0$ and $Y_1 = 0$. This solution is consitent with eq.(31) only if the limit $-\lambda_2/\lambda_1 = \infty$ is satisfied. As in the privious case, if $X_2 \to 0$ and $Y_2 \to 0$ the EoS $w_2 = (Y_2^2 - X_2^2)/(Y_2^2 + X_2^2) = (C^2-1)/(C^2+1)$ not determine and $-1 \le w_2 \le 1$. If $\lambda_2 \ne \infty$ we have $w_{2eff} = w_2$, while for $\lambda_2 = \infty$ we have $w_{2eff} \ge w_{1eff}$ and again the interaction term in eq.(28) does not need to vanish. In case 5 the value of the EoS for $Y_1$ is model dependent and must satisfy $w_{1eff} \ge w_{2eff}$.

| Num | $\lambda_1$ | $\lambda_2$ | $\lambda_3$ | $Y_1^2$ | $Y_2^2$ | $w_{1eff}$ | $w_{2eff}$ | $-\lambda_2/\lambda_1$ |
|---|---|---|---|---|---|---|---|---|
| 1 | $0$ | $\ne\infty$ | $0$ | $1$ | $0$ | $-1$ | $w_2$ | $\infty$ |
| 2 | $\lambda_1$ | $\infty$ | $0$ | $1$ | $0$ | $-1$ | $w_{2eff}$ | $\infty$ |
| 3 | $\lambda_1$ | $\lambda_2$ | $0$ | $1-\left(1+A^{2/3}\right)^{-1}$ | $\left(1+A^{2/3}\right)^{-1}$ | $-1$ | $-1$ | A |
| 4 | $\ne\infty$ | $0$ | $0$ | $0$ | $1$ | $-1$ | $-1$ | $0$ |
| 5 | $\infty$ | $\lambda_2$ | $0$ | $0$ | $1$ | $w_{1eff}$ | $-1$ | $0$ |
| 6 | $0$ | $\ne\infty$ | $\ne 0$ | $1$ | $0$ | $-1$ | $w_2$ | $\infty$ |
| 7 | $\lambda_1$ | $\infty$ | $\ne 0$ | $1$ | $0$ | $-1$ | $w_{2eff}$ | $\infty$ |

Table II. In this table we shows the different models that lead to an accelerating universe with $Y_1^2 + Y_2^2 = 1$ and when $\lambda_2/\lambda_1 < 0$. The values of $\lambda_1$, $\lambda_2$ and $\lambda_3$ are aribitray and can take any value (0, cte or $\infty$) but they need to satify the condition in the last column. The values of $w_{1eff}$, $w_{2eff}$ in models 2,5 and 8 are model dependent.

We now show one example of each case: $\lambda_3 = cte \ne 0$, $\lambda_3 = 0$ and $\lambda_3 = \infty$. We give in the appendix A a complete presentation of the differtent models that lead to an a accelerating universe at late times.

**Case 1** $\lambda_3 = cte \neq 0$.

**Model 1a:** $\lambda_1 = 0$, $\lambda_2 = \infty$, $\lambda_3 = cte \neq 0$.

The values of the $\lambda s$ coincide with case 7 in table II, with $Y_2^2 = 0$, $X_1 = X_2 = 0$ and the universe will be dominated at late time by a constant potential with $Y_1^2 = 1$, i.e. by the tachyon field with: $\Omega_1 = 1$, $\Omega_2 = 0$. From the eqs. (33) and (29) we get that $w_{1efec} = -1$, and the effective equation of state parameter by the field $\phi$ has to be in general $w_{2efec} \geq w_{1efec}$. The energy density of $\phi$ evolves slower than the tachyon field but we can still have $w_{1efec} \to -1$ and $w_{2efec} \to -1$. The case of $\lambda_3 \to cte$ corresponds to interaction potentials as $G(\phi) = V_0 e^{-c\phi + b/\phi}$ with $B(T,\phi) = F(T)G(\phi)$ and $F(T)$ arbitrary. From eq. (21) we obtain that $\lambda_3 = c + (b/\phi^2) \to c$ ($\phi \to \infty$, to minimize the potential $G(\phi)$). Since $Y_2^{\,2}=0$ we must have implies $\lambda_1/\lambda_2 \to 0$. Examples of potentiales leading to the values of the $\lambda s$, are for $\lambda_1 \to 0$ potentials such as $V(T)=K\exp(T^{-4})$, while for $\lambda_2 \to \infty$ requires potentials like $B=V_0 e^{-c\phi+b/\phi}T^3$. From eq. (35) the asymptotic relation between the scalar fields $T$ and $\phi$ is given in this case by $\phi=(-1/c)Ln\left(4K/3V_0T^7\right)$ where we observe that for $T\to\infty$ one has $\phi\to\infty$. From eqs.(21) we obtain that $\lambda_1 = 4/\sqrt{K}\,T^5\exp(T^{-4}/2)\to 0$ and $\lambda_2 = T\sqrt{27/4K\exp(T^{-4})}\to\infty$ as $T\to\infty$. As an example we present a model with the potentials $V(T)=\exp(T^{-4})$, $F(T)=T^3$, $G(\phi)=\exp(-\phi+1/\phi)$ and a barotropic fluid with $w_b=1/3$. In figs.1 we show the behavior of $\Omega_1$ and $\Omega_2$, the equation of state parameters $w_{1eff}\equiv w_{Teff}$ and $w_{2eff}\equiv w_{\phi eff}$ and the $\lambda_i$ parameters.

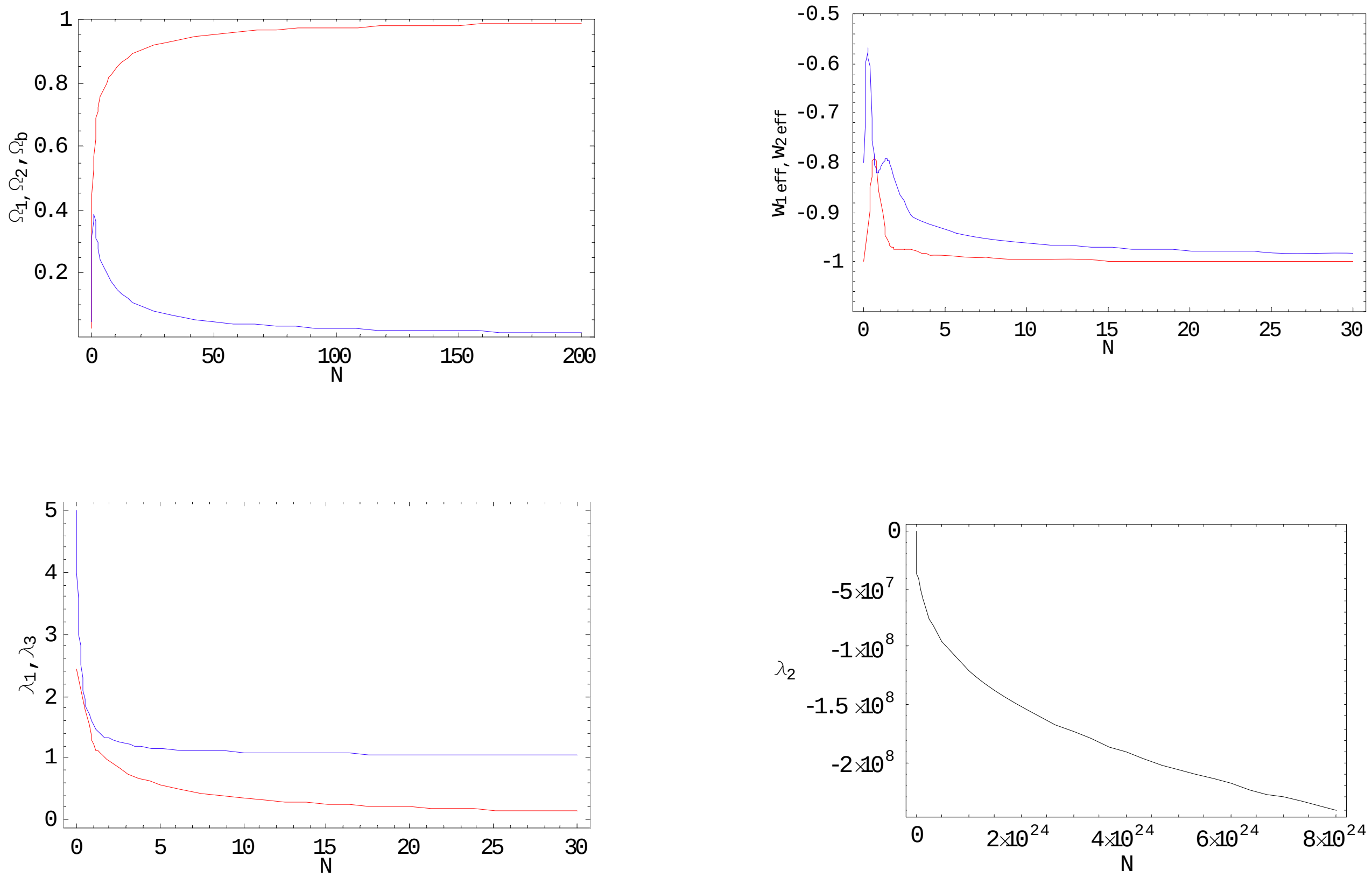

Fig 1. We show the evolution of $\Omega_1 = \Omega_T$ and $\Omega_2 = \Omega_\phi$ (gred and blue, respectively), the equation of state parameters $w_{1eff} = w_{Teff}$, $w_{2eff} = w_{\phi eff}$ (red and blue) where we appreciate that $w_{1eff}, w_{2eff} \to -1$ asymptotically with $w_{2eff} > w_{1eff}$, and the $\lambda_i$ parameters ($\lambda_1$ red, $\lambda_3$ blue) and $\lambda_2$ separately as a function of $N \equiv Lna$.

**Case 2** $\lambda_3=\infty$.

**Model 2a:** $\lambda_1=0$, $\lambda_2=\infty$, $\lambda_3=\infty$.

This model correspond to case 7 in Table II. In this limit we have again that $Y_2^{\,2}=0$ at late time (see eq. (32)). Since that $Y_1^2+Y_2^2=1$ we get $Y_1^2=1$. The universe is then dominated by the tachyon with: $\Omega_1=1$, $\Omega_2=0$ and from the eqs. (33) and (29) we get that $w_{1efec}=-1$. In fact in this case the effective equation of state parameter by the field $\phi$ has to be in general $w_{2efec}\ge w_{1efec}$. The case of $\lambda_3\to\infty$ corresponds to interaction potentials as $G(\phi)=V_0\,\mathrm{e}^{-\beta\phi^2}$ with $B(T,\phi)=F(T)G(\phi)$ and $F(T)$ arbitrary. From eq. (21) we obtain $\lambda_3=2\beta\phi\to\infty$ (as $\phi\to\infty$ on order to minimize the potential $G(\phi)$). The condition $Y_2^{\,2}=0$ implies that $\lambda_1/\lambda_2\to 0$. The limit of $\lambda_1\to 0$ includes potentials as $V(T)=K\exp(T^{-1})$, the case of $\lambda_2\to\infty$ require potentials such as $B=V_0e^{-\beta\phi^2}e^{\alpha T}$. From eq.(35) the asymptotic relation between the scalar fields $T$ and $\phi$ is given in this case by $\phi^2=(-1/\beta)Ln\left(K/\alpha V_0T^2e^{\alpha T}\right)$ where we observe that $T\to\infty$ for $\phi\to\infty$. From eqs.(21) we obtain that $\lambda_1=1/T^2\sqrt{K}\exp(T^{-1}/2)\to 0$ and $\lambda_2=T\sqrt{\alpha^3/K\exp(T^{-1})}\to T\sqrt{\alpha^3/K}\to\infty$ as $T\to\infty$. As example we present a model with the potentials $V(T)=(1/100)\exp(T^{-1})$, $F(T)=50T$, $G(\phi)=\exp(-100\phi^2)$ and a barotropic fluid with $w_b=1/3$. In fig.2 we show the behavior of $\Omega_1$, $\Omega_2$ and $\Omega_b$, the equation of state parameters $w_{1eff}\equiv w_{Teff}$ and $w_{2eff}\equiv w_{\phi eff}$ and the $\lambda_i$ parameters.

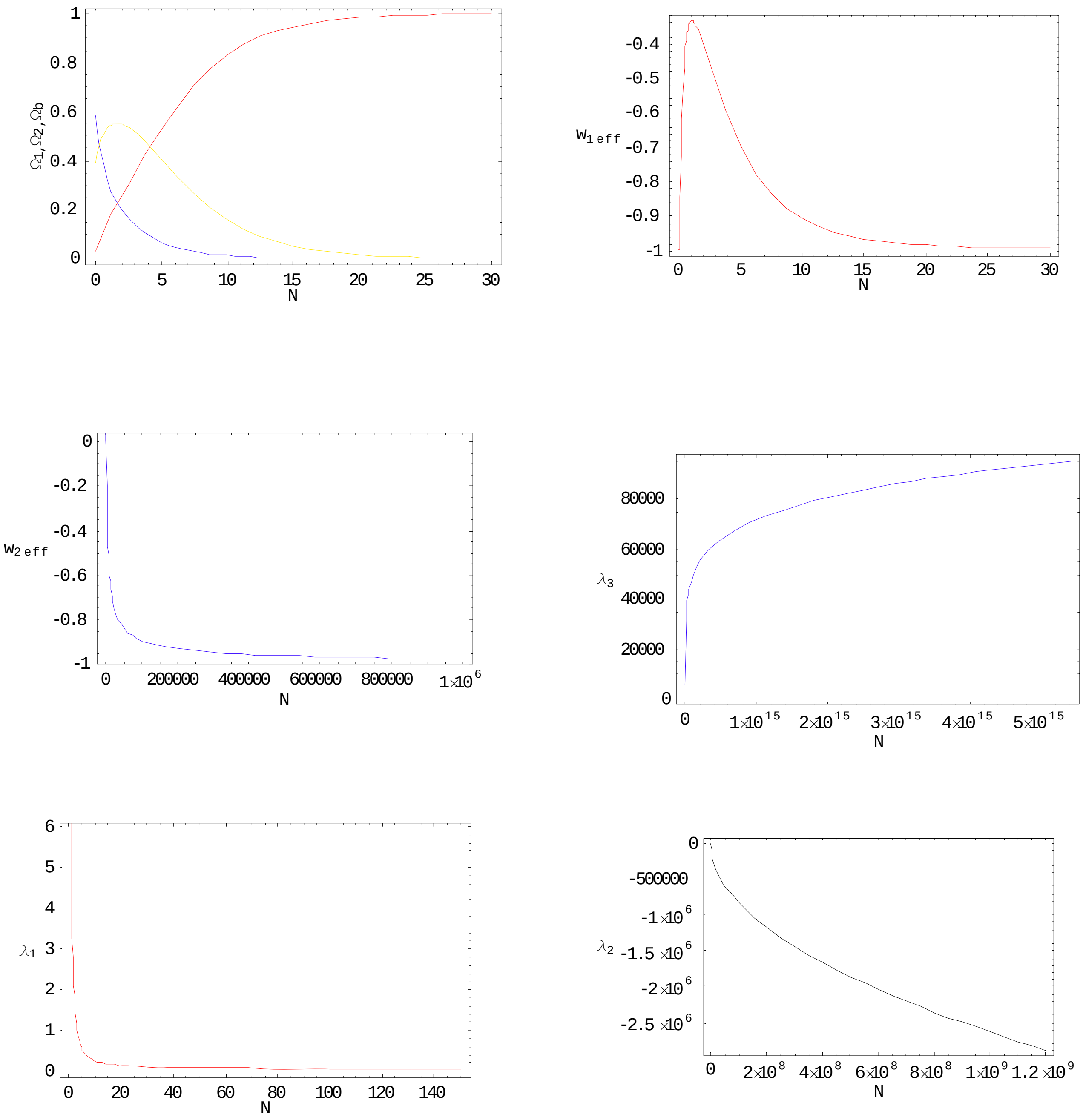

Fig 2. We show the evolution of $\Omega_1 = \Omega_T$, $\Omega_2 = \Omega_\phi$ and $\Omega_b$ (red, blue and yellow, respectively) the equation of state parameters $w_{1eff} = w_{Teff}$, $w_{2eff} = w_{\phi eff}$, where we see that $w_{1eff}, w_{2eff} \to -1$ asymptotically with $w_{2efec} \geq w_{1efec}$ and the parameters $\lambda_1$, $\lambda_2$ and $\lambda_3$ separately as a function of $N \equiv Lna$.

**Case 3** $\lambda_3 = 0$

**Model 3a:** $\lambda_1 = cte$, $\lambda_2 = cte$, $\lambda_3 = 0$.

This example corresponds to case 3 in Table II. Here both $Y_1$ and $Y_2$ can be different than zero as long with $Y_1^2 + Y_2^2 = 1$, the limit value of $\lambda_3 \to 0$ corresponds to interaction potentials as $G(\phi) = V_0 \phi^{-n}$ where $n > 0$ with $B(T,\phi) = F(T)G(\phi)$ and $F(T)$ arbitrary. From eq.(21) we obtain that $\lambda_3 = n/\phi \to 0$ ($\phi \to \infty$). If $Y_1^2 \neq 0$ and $Y_2^{\,2} \neq 0$ the effective equation of state parameters are $w_{1efec} = -1$ and $w_{2efec} = -1$ and we have A= $-\lambda_2/\lambda_1$ constant and $Y_1^{\ 2} = 1 - (1 + A^{2/3})^{-1}$, $Y_2^{\ 2} = (1 + A^{2/3})^{-1}$. The limit of $\lambda_1 \to cte$ includes potentials such as $V(T) = KT^{-2} e^{\beta/T}$, and the value of $\lambda_2 = cte \neq 0$ requires potentials as $B = V_0 \phi^{-n} T$ ($n > 0$). From eq.(35) the asymptotic relation between the scalar fields $T$ and $\phi$ is given in this case by $\phi^n = V_0 T^3 / 2K$ where we observe that $T \to \infty$ for $\phi \to \infty$. From eq.(21) we obtain that

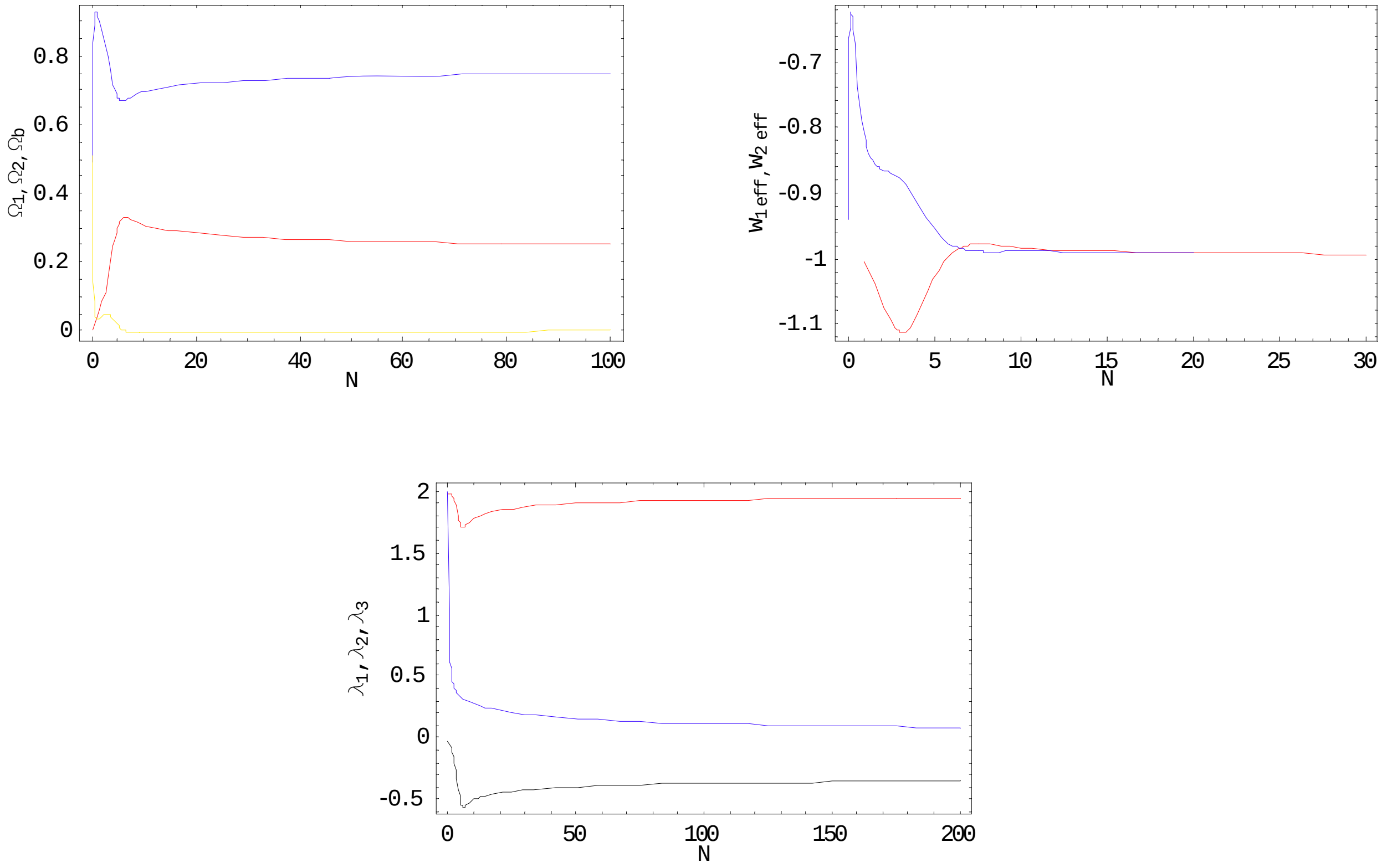

Fig3. We show the evolution of $\Omega_1 = \Omega_T$, $\Omega_2 = \Omega_\phi$ and $\Omega_b$ (red, blue and yellow, respectively) the equation of state parameters $w_{1eff} = w_{Teff}$, $w_{2eff} = w_{\phi eff}$ (red and blue) and the $\lambda_i$ parameters ($\lambda_1$ red, $\lambda_2$ balck and $\lambda_3$ blue ) as a function of $N \equiv Lna$.

$$\lambda_1 = 2 + \beta/T/\sqrt{K}\exp(\beta/2T) \to 2/\sqrt{K} \quad \text{and} \quad \lambda_2 = \sqrt{T}/\sqrt{Ke^{\beta/T}(b+2T)} \to 1/2\sqrt{K} \quad (T \to \infty).$$

As example we present a model with the potentials $V(T) = T^{-2} e^{-T^{-1}}$, $F(T) = 2T$ and $G(\phi) = \phi^{-1}$

and a barotropic fluid with $w_b = 1/3$. In fig.3 we show the behavior of $\Omega_1$, $\Omega_2$ and $\Omega_b$, the equation of state parameters $w_{1eff} \equiv w_{Teff}$ and $w_{2eff} \equiv w_{\phi eff}$ and the $\lambda_i$ parameters.

IV. SUMMARY AND CONCLUSIONS

We have studied the cosmological evolution of a universe filled with two scalar fields, a tachyon and a canonically normalized field, in the presence of a barotropic fluid in a FRW metric. We have shown that all model dependence is given in terms of three parameters, namely $\lambda_1 = -V_T/V$, $\lambda_2 = -B_T/B$ and $\lambda_3 = -B_\phi/B$, which in general are time dependent quantities. In a previous work [6] the parameters $\lambda_i's$ were assumed constant. Here our analysis is completely general and model independent. We do not assume any kind of tachyon potential $V(T)$ and we coupled $T$ to a canonical scalar field $\phi$ through an arbitrary but factorizeable interaction potential B (T,$\phi$) = F (T) $G$ ($\phi$). The solutions of the dynamical eqs.(14) through (17) are none linear and general analytic solutions do not exist. We determined the critical points for the cases where the universe ends accelerating and it give rise to twenty two possible models. To study them we find natural to separate the cases where asymptotically: $\lambda_3 \to cte \neq 0$, $\lambda_3 \to 0$ and $\lambda_3 \to \infty$. In all the models we gave the asymptotic relation between the scalar fields $T$ and $\phi$.

In the limit of $\lambda_3 \to cte$ we showed that $Y_2^2 = 0$ and from the condition $Y_1^2 + Y_2^2 = 1$, we deduced that $Y_1^2 = 1$. In this way we obtained a universe dominated by the tachyon with: $\Omega_1 = 1$, $\Omega_2 = 0$ with effective equation of state $w_{1efec} = -1$. The effective equation of state parameter by the fields must satisfy $w_{2efec} \geq w_{1efec}$ even though both $w$ may approach -1. In the case of $\lambda_3 \to \infty$ we obtained again that $Y_2^2 = 0$, $Y_1^2 = 1$ and the universe is dominated by the tachyon with $\Omega_1 = 1$, $\Omega_2 = 0$ and $w_{1efec} = -1$. The effective equation of state parameter of the field $\phi$ must be $w_{2efec} \geq w_{1efec}$, $\phi$ evolves slower than $T$, but we may have $w_{1efec} \to -1, w_{2efec} \to -1$. The case of $\lambda_3 \to \infty$ corresponds to interaction potentials such as $G(\phi) = V_0\, e^{-\beta\phi^2}$ with $B(T,\phi) = F(T)G(\phi)$ and $F(T)$ arbitrary. In the limit of $\lambda_3 \to 0$ we found that $Y_1$ and $Y_2$ are in general different of zero with $Y_1^2 + Y_2^2 = 1$. It means that we have a universe dominated by a constant potential due to both fields. From the solution $Y_1{}^2 = 1-(1+(-\lambda_2/\lambda_1)^{2/3})^{-1}$ and $Y_2{}^2 = (1+(-\lambda_2/\lambda_1)^{2/3})^{-1}$ we studied the different limits for the quotients $A$ = - $\lambda_2/\lambda_1$. Finally, models where $\lambda_2 \to 0$ asymptotically can be deduced from the combination of previous works (refs. [7] and [8]). In contrast, we determined here the cosmological evolution for models where the interaction between $\phi$ and $T$ does no vanish asymptotically leading to interesting cases where the universe accelerates at late times and may then be used to study dark energy.

Acknowledgements
This work was supported in part by CONACYT Project 80519 and IAC CONACYT Project.

Appendix A

We now present all the different cases that lead to an acceleraring universe with $Y_1^2 + Y_2^2 = 1$.

Model 1b: $\lambda_1 = cte \neq 0$, $\lambda_2 = \infty$, $\lambda_3 = cte \neq 0$

The limit of $\lambda_1 \to cte$ includes potentials as $V(T) = KT^{-2}e^{b/T}$ and the case of $\lambda_2 \to \infty$ requires potentials such as $B = V_0 e^{-c\phi + b/\phi} e^{\alpha T}$. From eq. (35) the asymptotic relation between the scalar fields $T$ and $\phi$ is given in this case by $\phi = (-1/c)Ln\left(2K/\alpha V_0 T^3 e^{\alpha T}\right)$ where we observe that $T \to \infty$ for $\phi \to \infty$. From eqs.(21) we obtain that $\lambda_1 = 2 + b/T/\sqrt{K}\exp(b/2T) \to 2/\sqrt{K} = cte$ and $\lambda_2 = T^2\sqrt{\alpha^3/(b+2T)K\,\mathrm{e}^{b/T}} \to \infty$ ($T \to \infty$).

Model 1c: $\lambda_1 = \infty$, $\lambda_2 = \infty$, $\lambda_3 = cte \neq 0$

This model require that $\lambda_1 \to \infty$ slower than $\lambda_2 \to \infty$ to have $\lambda_1/\lambda_2 \to 0$, the limit of $\lambda_1 \to \infty$ includes potentials like $V(T) = KT^{-4}$ and the case of $\lambda_2 \to \infty$ requires potentials as $B = V_0 e^{-c\phi + b/\phi} e^{\alpha T}$. From eq. (35) the asymptotic relation between the scalar fields $T$ and $\phi$ is given in this case by $\phi = (-1/c)Ln\left(4K/\alpha V_0 e^{\alpha T} T^5\right)$ where we observe that $T \to \infty$ for $\phi \to \infty$. From eqs. (21) we obtain that $\lambda_1 = 4T/\sqrt{K} \to \infty$ and $\lambda_2 = \sqrt{\alpha^3 T^5/4K} \to \infty$ (in the limit of $T \to \infty$) in such way that $\lambda_1/\lambda_2 = 8/\alpha^{3/2}T^{3/2} \to 0$.

Model 1d: $\lambda_1 = 0$, $\lambda_2 = 0$, $\lambda_3 = cte \neq 0$.

Since $\lambda_2 \to 0$ this model is reduced at late time to two uncoupled scalar fields acting independently. From ref. [8] we know that this model does accelerate, independently of $\phi$, because the tachyon potentials with $\lambda_1 \to 0$ always lead to an accelerating universe. It is relevant to note that $T$ ends dominating the universe independently from the value of the constant $c$ in the canonical scalar potential $V_0 e^{-c\phi + b/\phi}$ (where $\lambda_3$ reaches $c$ asymptotically). This model requires that $\lambda_1 \to 0$ faster than $\lambda_2 \to 0$ to have $\lambda_1/\lambda_2 \to 0$. The limit of $\lambda_1 \to 0$ includes potentials such as $V(T) = K\exp(T^{-1})$, the limit value of $\lambda_2 = 0$ require potentials like $B = V_0 e^{-c\phi + b/\phi} T^2$. From eq.(35) the asymptotic relation between the scalar fields $T$ and $\phi$ is given in this case by $\phi = (-1/c)Ln\left(K/2V_0 T^3\right)$ where we observe that $T \to \infty$ for $\phi \to \infty$. From eqs.(21) we obtain that $\lambda_1 = 1/T^2\sqrt{K}\exp(T^{-1}/2) \to 0$ and $\lambda_2 = \sqrt{8/KT\exp(T^{-1})} \to \sqrt{8/KT} \to 0$ (as $T \to \infty$) with $\lambda_1/\lambda_2 = 1/\sqrt{8}T^{3/2} \to 0$.

Model 1e: $\lambda_1 = 0$, $\lambda_2 = cte \neq 0$, $\lambda_3 = cte \neq 0$.

The limit of $\lambda_1 \to 0$ includes potentials as $V(T) = K\exp(T^{-1})$, and the case in which $\lambda_2 = cte \neq 0$ require potentials as $B = V_0 e^{-c\phi + b/\phi} e^{T^{1/3}}$. From eq.(35) the asymptotic relation between the scalar fields $T$ and $\phi$ is given in this case by $\phi=(-1/c)Ln\left(3K/V_0e^{T^{1/3}}T^{4/3}\right)$ where we observe again that $T\to\infty$ for $\phi\to\infty$. From eqs.(21) we obtain that $\lambda_1=1/T^2\sqrt{K}\exp(T^{-1}/2)\to 0$ and $\lambda_2=\sqrt{1/27K\exp(T^{-1})}\to\sqrt{1/27K}=cte$ (as $T\to\infty$).

Model 2b: $\lambda_1=cte\neq 0$, $\lambda_2=\infty$, $\lambda_3=\infty$.

The limit of $\lambda_1\to cte$ includes potentials as $V(T)=KT^{-2}e^{b/T}$, the case of $\lambda_2\to\infty$ require potentials as $B=V_0e^{-\beta\phi^2}e^{\alpha T}$. From eq. (35) the asymptotic relation between the scalar fields $T$ and $\phi$ is given in this case by $\phi^2=(-1/\beta)Ln\left(2K/\alpha V_0T^3e^{\alpha T}\right)$ where we observe that $T\to\infty$ for $\phi\to\infty$. From eqs.(21) we obtain that $\lambda_1=2+b/T\big/\sqrt{K}\exp(b/2T)\to 2/\sqrt{K}=cte$ and $\lambda_2=T^2\sqrt{\alpha^3/(b+2T)K\,\mathrm{e}^{b/T}}\to\infty$ ($T\to\infty$).

Model 2c: $\lambda_1=\infty$, $\lambda_2=\infty$, $\lambda_3=\infty$.

This model requires that $\lambda_1\to\infty$ slower than $\lambda_2\to\infty$ to have $\lambda_1/\lambda_2\to 0$. The limit of $\lambda_1\to\infty$ includes potentials as $V(T)=KT^{-4}$ and the case of $\lambda_2\to\infty$ requires for example potentials such as $B=V_0e^{-\beta\phi^2}e^{\alpha T}$. From eq.(35) the asymptotic relation between the scalar fields $T$ and $\phi$ is given in this case by $\phi^2=(-1/\beta)Ln\left(4K/\alpha V_0e^{\alpha T}T^5\right)$ where we observe that $T\to\infty$ for $\phi\to\infty$. From eq.(21) we obtain that $\lambda_1=4T/\sqrt{K}\to\infty$ and $\lambda_2=\sqrt{\alpha^3T^5/4K}\to\infty$ (in the limit of $T\to\infty$) in such a way that $\lambda_1/\lambda_2=8/\alpha^{3/2}T^{3/2}\to 0$.

Model 2d: $\lambda_1=0$, $\lambda_2=0$, $\lambda_3=\infty$.

Since in the limit $\lambda_2\to 0$ the fields $T$ and $\phi$ end up uncoupled, from $\lambda_1\to 0$ we can see that we will get an accelerating universe dominated by the tachyon with $w_{1efec}=-1$ (see ref [8]), because the case of $\lambda_3\to\infty$ never leads to $\phi$ to dominate the universe with an acceleration (see ref. [7] or section IIA). This model requires that $\lambda_1\to 0$ faster than $\lambda_2\to 0$ to have $\lambda_1/\lambda_2\to 0$, the limit of $\lambda_1\to 0$ includes potentials as $V(T)=K\exp(T^{-1})$, the limit value of $\lambda_2=0$ requires potentials as $B=V_0e^{-\beta\phi^2}e^{-T^{-1/3}}$. From eq.(35) the asymptotic relation between the scalar fields $T$ and $\phi$ is given in this case by $\phi^2=(-1/\beta)Ln\left(3K/V_0T^{2/3}\right)$ where we observe that $T\to\infty$ for $\phi\to\infty$. From eqs.(21) we obtain that $\lambda_1=1/\sqrt{K}\,T^2\exp(T^{-1}/2)\to 0$ and $\lambda_2=1/T\sqrt{27K\exp(T^{-1})}\to 1/T\sqrt{27K}\to 0$, in the limit of $T\to\infty$ in such way that $\lambda_1/\lambda_2=\sqrt{27}/T\to 0$.

Model 3b: $\lambda_1=\infty$, $\lambda_2=0$, $\lambda_3=0$.

Since $\lambda_2\to 0$ this model is reduced at late times to two uncoupled scalar fields acting separately. From the combination of the works given in [7] and [8], we can to predict that $\phi$ will dominate since the potentials as $G(\phi)=V_0\phi^{-n}$ always leads to an accelerating universe dominated by $\phi$ with $w_{2efec}=-1$, as tachyon potentials consistent with $\lambda_1\to\infty$ end with $w_{1efec}>-1$ (see section IIA). Since $\lambda_2/\lambda_1\to 0$ we get that $Y_1^2=0$ and $Y_2^2=1$, at late time and we have a universe dominated by the canonical scalar field with $\Omega_2=1, \Omega_1=0$ and the effective equation of state parameter $w_{2efec}=-1$ (see eq. (28)). The limit of $\lambda_1\to\infty$ includes potentials as $V(T)=KT^{-3}$, the limit value of $\lambda_2=0$ requires potentials as $B=V_0\phi^{-n}\exp[-e^{-\beta T}]$ ($n>0$). From eq.(35) the asymptotic relation between the scalar fields $T$ and $\phi$ is given in this case by $\phi^n=\beta V_0 T^4/3Ke^{\beta T}$ where we observe that $T\to\infty$ for $\phi\to\infty$. From eqs.(21) we obtain that $\lambda_1=3\sqrt{T}/\sqrt{K}\to\infty$ and $\lambda_2=\sqrt{\beta^3/3K}\left(T^2/\exp(3\beta T/2)\right)\to 0$ (in the limit of $T\to\infty$).

Model 3c: $\lambda_1=\infty$, $\lambda_2=cte\neq 0$, $\lambda_3=0$.

Since $\lambda_2/\lambda_1\to 0$ we get that $Y_1^2=0$ and $Y_2^2=1$, at late times and we have a universe dominated by the canonical scalar field with $\Omega_2=1, \Omega_1=0$ and the effective equation of state parameter $w_{2efec}=-1$. In this model the effective equation of state parameter by the tachyon has to be in general $w_{1efec}\geq w_{2efec}$ but both w may go to -1. The limit of $\lambda_1\to\infty$ includes potentials as $V(T)=KT^{-3}$ and the limit value of $\lambda_2=cte\neq 0$ requires potentials as $B=V_0\phi^{-n}e^{-\beta T^{-1/3}}$ (with $n>0$). From eq.(35) the asymptotic relation between the scalar fields $T$ and $\phi$ is given in this case by $\phi^n=\beta V_0 T^{8/3}/9K$ where we observe that $T\to\infty$ for $\phi\to\infty$. From eqs.(21) we obtain that $\lambda_1=\left(3/\sqrt{K}\right)\sqrt{T}\to\infty$ and $\lambda_2=\beta^{3/2}/T\sqrt{27K\exp(T^{-1})}\to\beta^{3/2}/T\sqrt{27K}\to 0$ in the limit of $T\to\infty$.

Model 3d: $\lambda_1=\infty$, $\lambda_2=\infty$, $\lambda_3=0$.

The value of the quotient $\lambda_1/\lambda_2$, in the limit of $\lambda_1\to\infty$ and $\lambda_2\to\infty$ determines the asymptotic value of the quantities $Y_1$ and $Y_2$. For $\lambda_1\to\infty$ and $\lambda_2\to\infty$ at the same rate, then $\lambda_1/\lambda_2\to A\neq 0$, and we simply have $Y_1^2=\left(1+A^{2/3}\right)^{-1}$ and $Y_2^2=1-\left(1+A^{2/3}\right)^{-1}$ with $w_{1efec}=-1$ (from eq. (29) we require that $\lambda_1 Y_1 X_1=0$) and $w_{2efec}=-1$ (see eq. (28)) in such a way that $w_{1efec}/w_{2efec}=1$ asymptotically. The universe is dominated at late time for a constant potential which depends on $T$ and $\phi$. The limit of $\lambda_1\to\infty$ includes potentials as $V(T)=K/T^4$, the limit value of $\lambda_2=\infty$ require potentials as $B=V_0\phi^{-n}T$ ($n>0$). From eq.(35) the asymptotic relation between the scalar fields $T$ and $\phi$ is given in this case by $\phi^n=V_0T^5/4K$ where we observe that $T\to\infty$ for $\phi\to\infty$. From eqs.(21) we obtain that $\lambda_1=4T/\sqrt{K}\to\infty$ and $\lambda_2=T/\sqrt{K}\to\infty$ ($T\to\infty$) in such a way that $\lambda_1/\lambda_2=4$.

If $\lambda_1\to\infty$ slower than $\lambda_2\to\infty$ then $\lambda_1/\lambda_2\to 0$ and $Y^2{}_1\to 1$ $Y^2{}_2\to 0$ with $w_{1efec}=-1$ (from eq. (29) we require that $\lambda_1 Y_1 X_1=0$) and in general $w_{2efec}\geq w_{1efec}$. In this way the universe is dominated at late time for a constant potential which depends on $T$ with the values of $\Omega_1 = 1, \Omega_2 = 0$. The limit of $\lambda_1 \to \infty$ includes potentials such as $V(T) = KT^{-3}$, the limit value of $\lambda_2 = \infty$ requires potentials as $B = V_0\phi^{-n}e^{\beta T}$ ($n > 0$). From eq. (35) the asymptotic relation between the scalar fields $T$ and $\phi$ is given in this case by $\phi^n = V_0\beta e^{\beta T}T^4/3K$ where we observe that $T \to \infty$ for $\phi \to \infty$. From eqs.(21) we obtain that $\lambda_1 = 3T^{1/2}/\sqrt{K} \to \infty$ and $\lambda_2 = T^2\sqrt{\beta^3/3K} \to \infty$ ($T \to \infty$) in such a way that $\lambda_1/\lambda_2 = 3\sqrt{3}/\beta^{3/2}T^{3/2} \to 0$.

If $\lambda_1 \to \infty$ faster than $\lambda_2 \to \infty$ then $\lambda_2/\lambda_1 \to 0$ we get that $Y_1^2 = 0$ and $Y_2^2 = 1$, at late time and we have a universe dominated by the scalar field canonical with: $\Omega_2 = 1, \Omega_1 = 0$ with the effective equation of state parameter $w_{2efec} = -1$. In fact in this model the effective equation of state parameter by the tachyon has to be in general $w_{1efec} \ge w_{2efec}$ (and $\lambda_1 Y_1 X_1/\sqrt{3} \ge 0$ (see eq. (29)). The limit of $\lambda_1 \to \infty$ includes potentials like $V(T) = KT^{-6}$, the limit value of $\lambda_2 = \infty$ require potentials as $B = V_0\phi^{-n}e^{-T^{-1}}$ ($n > 0$). From eq.(35) the asymptotic relation between the scalar fields $T$ and $\phi$ is given in this case by $\phi^n = V_0T^5/6K$ where we observe that $T \to \infty$ for $\phi \to \infty$. From eqs.(21) we obtain that $\lambda_1 = 6T^2/\sqrt{K} \to \infty$ and $\lambda_2 = \sqrt{T/6K} \to \infty$ ($T \to \infty$) in such a way that $\lambda_2/\lambda_1 = 1/6\sqrt{6}T^{3/2} \to 0$.

Model 3e: $\lambda_1 = 0, \lambda_2 = \infty$, $\lambda_3 = 0$.

Since $\lambda_1/\lambda_2 \to 0$ we have then $Y^2{}_1 \to 1$, $Y^2{}_2 \to 0$ with $w_{1efec} = -1$ (see eq. (29)) and in general $w_{2efec} \ge w_{1efec}$. In this model the universe is dominated at late time by a constant potential which depends on $T$ with $\Omega_1 = 1, \Omega_2 = 0$. The limit of $\lambda_1 \to 0$ includes potentials as $V(T) = K\exp(T^{-1})$, the limit value of $\lambda_2 = \infty$ requires potentials as $B = V_0\phi^{-n}e^{\alpha T}$ ($n > 0$). From eq.(35) the asymptotic relation between the scalar fields $T$ and $\phi$ is given in this case by $\phi^n = \alpha V_0 T^2 e^{\beta T}/K$ where we observe that $T \to \infty$ for $\phi \to \infty$. From eqs.(21) we obtain that $\lambda_1 = 1/T^2\sqrt{K}\exp(T^{-1}/2) \to 0$ and $\lambda_2 = T\sqrt{\alpha^3/K\exp(T^{-1})} \to T\sqrt{\alpha^3/K} \to \infty$ (as $T \to \infty$).

Model 3f: $\lambda_1 = cte$, $\lambda_2 = \infty$, $\lambda_3 = 0$.

Since $\lambda_1/\lambda_2 \to 0$ then $Y^2{}_1 \to 1$ $Y^2{}_2 \to 0$ with $w_{1efec} = -1$(see eq. (29)) and $w_{2efec} \ge -1$. In this model the universe is dominated at late time by a constant potential which depends on $T$ with the values of $\Omega_1 = 1, \Omega_2 = 0$. The case of $\lambda_1 \to cte$ includes potentials as $V(T) = KT^{-2}e^{b/T}$ and $\lambda_2 = \infty$ requires potentials as $B = V_0\phi^{-n}e^{\beta T}$ ($n > 0$). From eq. 35) the asymptotic relation between the scalar fields $T$ and $\phi$ is given in this case by $\phi^n = \beta V_0 T^3 e^{\beta T}/2K$ where we observe that $T \to \infty$ for $\phi \to \infty$. From eqs. (21) we obtain that $\lambda_1 = 2 + b/T/\sqrt{K}\exp(b/2T) \to 2/\sqrt{K} = cte$ and $\lambda_2 = T^2\sqrt{\alpha^3/(b+2T)K\,\mathrm{e}^{b/T}} \to \infty$ ($T \to \infty$).

Model 3g: $\lambda_1 = 0, \lambda_2 = cte, \lambda_3 = 0$.

In this case we have $\lambda_1/\lambda_2 = 0$ giving $Y_1^2 = 1$ and $Y_2^2 = 0$ at late times. Clearly the universe is dominated at late time by the tachyon with $\Omega_1 = 1,\ \Omega_2 = 0$. $T$ has an effective equation of state parameter $w_{1efec} = -1$ (see eq. (29)) and in general $\phi$ has an $w_{2efec} \geq w_{1efec}$. The limit of $\lambda_1 \to 0$ includes potentials as $V(T) = K\exp(T^{-1})$, the limit value of $\lambda_2 = cte \neq 0$ requires potentials as $B = V_0\phi^{-n}e^{T^{1/3}}$ ($n > 0$). From eq. (35) the asymptotic relation between the scalar fields $T$ and $\phi$ is given in this case by $\phi^n = V_0 T^{4/3} e^{T^{1/3}} \big/ 3K$ where we observe that $T \to \infty$ for $\phi \to \infty$. From eqs.(21) we obtain that $\lambda_1 = 1\big/\sqrt{K}\,T^2 \exp(T^{-1}/2) \to 0$ and $\lambda_2 = 1\big/\sqrt{27K\exp(T^{-1})} \to 1\big/\sqrt{27K} = cte$ as $T \to \infty$.

Model 3h: $\lambda_1 = cte$, $\lambda_2 = 0$, $\lambda_3 = 0$.

Since in the limit $\lambda_2 \to 0$ the fields $T$ and $\phi$ end up uncoupled we know that we will get an accelerating universe (independently of $T$) because the potentials as $G(\phi) = V_0\phi^{-n}$ always lead to an accelerating universe dominated by $\phi$ with $w_{2efec} = -1$ (see section II A and ref 7). Is relevant to note (as we will see) that $\phi$ end dominating the universe independently from the constant value of $\lambda_1 = cte$ (see section II A and ref. [8]).

In this case we obtain that $\lambda_2/\lambda_1 \to 0$ and $Y_1^2 = 0$ and $Y_2^2 = 1$ at late time. So the universe is dominated at late time by the canonical scalar field with $\Omega_2 = 1,\ \Omega_1 = 0$ with the effective equation of state parameter $w_{2efec} = -1$. In fact in this model the effective equation of state parameter by the tachyon is too $w_{1efec} = -1$ (see eq. (29)). In this case the field $\phi$ evolves faster than the tachyon field does it, so $w_{1efec} > w_{2efec}$ although $w_{1efec} \to -1,\ w_{2efec} \to -1$. The limit of $\lambda_1 \to cte$ includes potentials as $V(T) = KT^{-2}e^{\beta/T}$, the limit value of $\lambda_2 = 0$ requires potentials as $B = V_0\phi^{-n}e^{-T^{-1}}$ ($n > 0$). From eq. (35) the asymptotic relation between the scalar fields $T$ and $\phi$ is given in this case by $\phi^n = V_0 T/2K \to \infty$ where we observe that $T \to \infty$ for $\phi \to \infty$. From eqs.(21) we obtain that $\lambda_1 = 2 + \beta/T\big/\sqrt{K}\exp(\beta/2T) \to 2\big/\sqrt{K} = cte$ and $\lambda_2 = 1\big/T\sqrt{e^{\beta/T}K(2T+\beta)} \to 1\big/\sqrt{2KT^3} \to 0$ ($T \to \infty$).

Model 3i: $\lambda_1 = 0$, $\lambda_2 = 0$, $\lambda_3 = 0$.

Since $\lambda_2 \to 0$ this model is reduced at late time to two uncoupled scalar fields acting separately. From ref. [7] we know that this model always accelerate (independently of $T$) because the potentials as $G(\phi) = V_0\phi^{-n}$ leads to an accelerating universe dominated by $\phi$ with $w_{2efec} = -1$. Since the limit of $\lambda_1 \to 0$ is consistent with accelerating models dominated by the tachyon with $w_{1efec} = -1$ (see [8]) (independently of $\phi$), the question of which of the scalar fields end dominating at late time depend heavily from the interaction term, such as we will see next.

We classify this model in according with the asymptotic value of the quotient $\lambda_1/\lambda_2$ in the limit of $\lambda_1 \to 0$ and $\lambda_2 \to 0$ shown in the following three cases:

i) $\lambda_1 \to 0$, $\lambda_2 \to 0$ and $\lambda_1/\lambda_2 \to cte = A \neq 0$.

In this case we simply have $Y_1^2 = \left(1+A^{2/3}\right)^{-1}$ and $Y_2^2 = 1-\left(1+A^{2/3}\right)^{-1}$ with $w_{1efec} = -1$ and $w_{2efec} = -1$ (see eqs. (28) and (29)) in such a way that $w_{1efec}/w_{2efec} = 1$ asymptotically. So the universe is dominated at late time by a constant potential which depends on $T$ and $\phi$. The limit of $\lambda_1 \to 0$ includes potentials as $V(T) = KT^{-1}$, the limit value of $\lambda_2 = 0$ requires potentials as $B = V_0\phi^{-n}T$ ($n > 0$). From eq.(35) the asymptotic relation between the scalar fields $T$ and $\phi$ is given in this case by $\phi^n = V_0 T^2/K$ where we observe that $T \to \infty$ for $\phi \to \infty$. From eqs.(21) we obtain that $\lambda_1 = 1/T^{1/2}\sqrt{K} \to 0$ and $\lambda_2 = 1/T^{1/2}\sqrt{K} \to 0$ ($T \to \infty$) in such a way that $\lambda_2/\lambda_1 = 1$.

ii) $\lambda_1 \to 0$, $\lambda_2 \to 0$ with $\lambda_1/\lambda_2 \to 0$.

If $\lambda_1$ approaches zero faster than $\lambda_2$ then we have $\lambda_1/\lambda_2 \to 0$ and $Y^2{}_1 \to 1, Y^2{}_2 \to 0$. As consequence $\Omega_1 = 1$, $\Omega_2 = 0$ and the universe end accelerating and dominated at late time by the tachyon field with the effective equation of state parameter $w_{1efec} = -1$. In this case the effective equation of state parameter for $\phi$ has to be in general $w_{2efec} \geq w_{1efec}$. The limit of $\lambda_1 \to 0$ includes potentials as $V(T) = K\exp(T^{-1})$, the limit value of $\lambda_2 = 0$ require potentials as $B = V_0\phi^{-n}e^{-T^{-1/3}}$ ($n > 0$). From eq.(35) the asymptotic relation between the scalar fields $T$ and $\phi$ is given in this case by $\phi^n = V_0 T^{2/3}/3K$ where we observe that $T \to \infty$ for $\phi \to \infty$. From eqs. (21) we obtain that $\lambda_1 = 1/T^2\sqrt{K}\exp(T^{-1}/2) \to 0$ and $\lambda_2 = 1/T\sqrt{27K\exp(T^{-1})} \to 1/T\sqrt{27K} \to 0$ ($T \to \infty$) in such a way that $\lambda_1/\lambda_2 = \sqrt{27}/T \to 0$.

iii) $\lambda_1 \to 0$, $\lambda_2 \to 0$ with $\lambda_2/\lambda_1 \to 0$.

If $\lambda_1$ approaches zero slower than $\lambda_2$ then we have $\lambda_2/\lambda_1 \to 0$ $Y_1^2 = 0$, $Y_2{}^2 = 1$ and we have a universe dominated at late time by the scalar field canonical with $\Omega_2 = 1$, $\Omega_1 = 0$ with the effective equation of state parameter $w_{2efec} = -1$. In fact in this model the effective equation of state parameter by the tachyon is too $w_{1efec} = -1$ (see eq. (29)). It indicates that the field $\phi$ evolves faster than the tachyon field does so $w_{1efec} > w_{2efec}$ although $w_{1efec} \to -1$ $w_{2efec} \to -1$. The limit of $\lambda_1 \to 0$ includes potentials as $V(T) = KT^{-3/2}$, the limit value of $\lambda_2 = 0$ require potentials as $B = V_0\phi^{-n}e^{-T^{-1/2}}$ ($n > 0$). From eq. (35) the asymptotic relation between the scalar fields $T$ and $\phi$ is given in this case by $\phi^n = V_0 T/3K \to \infty$, where we observe that $T \to \infty$ for $\phi \to \infty$. From eq.(21) we obtain that $\lambda_1 = \left(3/2\sqrt{K}\right)1/T^{1/4} \to 0$ and $\lambda_2 = 1/T\sqrt{12K} \to 0$ ($T \to \infty$) in such a way that $\lambda_2/\lambda_1 = 1/T^{3/4}\sqrt{27} \to 0$.